\documentclass[aps,prl,twocolumn]{revtex4}
\usepackage[dvips]{graphicx}

\begin{document}
\title{
Asperity characteristics of the Olami-Feder-Christensen model of earthquakes
}

\author{
 Hikaru Kawamura, Takumi Yamamoto, Takeshi Kotani and Hajime Yoshino
}

\affiliation{Department of Earth and Space Science, Faculty of Science,
Osaka University, Toyonaka 560-0043,
Japan}
\date{\today}
\begin{abstract}
Properties of the Olami-Feder-Christensen (OFC) model of earthquakes are studied by numerical simulations. The previous study indicated that the model exhibited ``asperity''-like phenomena, {\it i.e.\/}, the same region ruptures  many times near periodically [T.Kotani {\it et al\/}, Phys. Rev. E {\bf 77}, 010102 (2008)]. Such periodic or characteristic features apparently coexist with power-law-like critical features, {\it e.g.\/}, the Gutenberg-Richter law observed in the size distribution.  In order to clarify the origin and the nature of the asperity-like phenomena, we investigate here the properties of the OFC model with emphasis on its stress distribution. It is found that the asperity formation is accompanied by self-organization of the highly concentrated stress state. Such stress organization naturally provides the mechanism underlying our observation that a series of asperity events repeat  with a common epicenter site and with a common  period solely determined by the transmission parameter of the model. Asperity events tend to cluster both in time and in space.
\end{abstract}
\maketitle

\section{Introduction}

 Statistical properties of earthquakes have attracted much interest in seismology as well as in  statistical physics \cite{Scholzbook,Kolkotta}. Statistical properties of earthquakes are often characterized by power-laws,  {\it e.g.\/}, the Gutenberg-Richter (GR) law observed in earthquake size distribution, or the Omori law observed in the time evolution of aftershocks frequency.  The concept of ``self-organized criticality'' (SOC) was advocated by P. Bak \cite{Bak}. In this view, the Earth's crust is supposed to be in a critical state, while seismicity has a close analogy to critical phenomena of thermodynamic second-order transition.

 Several statistical models of earthquakes embodying such SOC features of earthquakes have been proposed and studied. One standard model is the so-called spring-block or the Burridge-Knopoff (BK) model, in which an earthquake fault is modeled as an assembly of blocks mutually connected via elastic springs which are slowly driven by external force \cite{BK,CL89,CL94}. While this model sometimes exhibits critical or near-critical properties under restricted conditions, generic properties of this model are not necessarily critical, often exhibiting ``characteristic'' properties accompanied by characteristic energy and time scales. For example, the BK model with the nearest-neighbor \cite{MoriKawa1DSR,MoriKawa2DSR} or the long-range interaction \cite{MoriKawaLR} has turned out to exhibit either ``supercritical'', ``subcritical'' or ``near-critical'' behavior depending on the parameters of the model. 

 Another statistical model extensively studied in statistical physics in the context of SOC might be the so-called OFC model, which was first introduced by Olami, Feder and Christensen (OFC) as a further simplification of the BK model \cite{OFC}. It is a two-dimensional lattice model where the rupture propagates from lattice site to its nearest-neighbor sites in a non-conservative manner, often causing multisite seismic events or ``avalanches''. Numerical studies have revealed that the OFC model exhibits apparently critical properties such as the GR law \cite{OFC,Grassberger,Lise} or the  Omori law \cite{Hergarten}, although there still remains controversy concerning whether the behavior of this model is strictly critical \cite{Lise} or only approximately so \cite{Prado,Boulter,Drossel}. In any case, the OFC model has been regarded as a typical non-conservative model exhibiting SOC.

 Earthquakes in nature sometimes exhibit characteristic properties accompanied by characteristic energy and time scales \cite{Scholzbook,Kolkotta}. In contrast to its critical feature, possible ``characteristic'' feature of the OFC model, if any, has attracted much less attention so far. Botani and Delamote observed that synchronized clusters of various sizes existed in the steady state of the OFC model with a large, but finite lifetime \cite{Bottani}. More recently, Kotani, Yoshino and Kawamura have revealed strikingly characteristic features of the model, by demonstrating that the local recurrence-time distribution of the OFC model exhibits a sharp $\delta$-function-like peak corresponding to rhythmic recurrence of events with a fixed period uniquely determined by the transmission parameter of the model \cite{Kotani}, together with a power-law-like tail corresponding to scale-free recurrence of events. In fact, the OFC model was found to exhibit phenomena closely resembling the ``asperity'' known in seismology. Such an asperity repeats to rupture sequentially, often  more than ten times, with almost the same period and with the same site as an epicenter (triggering site) \cite{Kotani}.  Statistics of epicenter sequences of the model was also studied by Peixoto and Prado in the framework of a growing complex network \cite{Peixoto}.

 In the present paper, following Ref.\cite{Kotani}, we further investigate spatiotemporal correlations of the OFC model, with particular emphasis on its characteristic features. In particular, we wish to unravel more detailed properties of the ``asperity'' involved in the model.  It is found that, in the formation of the asperity, self-organization process occurs in which the stress in the asperity region gets more and more ``quantized'' to discretized values. This self-organization process promotes an epicenter of the preceding event to become an epicenter of the next event again with the period uniquely determined by the transmission parameter of the model. It is also observed that the epicenter of large events tends to lie at the tip (or at the corner) of the rupture zone, rather than in its interior.

After defining the model and explaining some of its basic properties in \S 2, we present our numerical data of the local recurrence-time distribution and of the stress distribution in \S 3. Various asperity characteristics are investigated in detail in  \S 4. The relation between the self-organization of the highly concentrated stress state and the asperity characteristics are clarified. Section 5 is devoted to summary and discussion. Analysis of the time evolution of the stress concentration is given in Appendix.

\section{The Model}

 In the OFC model,  ``stress'' variable $f_i$ ($f_i\geq 0$) is assigned to each site on a square lattice with $L\times L$ sites. Initially,  a random value in the interval [0,1] is assigned to each $f_i$, while $f_i$ is increased with a constant rate uniformly over the lattice until, at a certain site $i$, the $f_i$ value reaches a threshold, $f_c=1$. Then, the site $i$ ``topples'', and a fraction of stress $\alpha f_i$ ($0<\alpha<0.25$) is transmitted to its four nearest neighbors, while $f_i$ itself is reset to zero.  If the stress of some of the neighboring sites $j$ exceeds the threshold, {\it i.e.\/}, $f_j\geq f_c=1$, the site $j$ also topples, distributing a fraction of stress $\alpha f_j$ to its four nearest neighbors. Such a sequence of topplings continues until the stress of all sites on the lattice becomes smaller than the threshold $f_c$. A sequence of toppling events, which is assumed to occur instantaneously, corresponds to one seismic event or an ``avalanche''. After an avalanche, the system goes into an interseismic period where uniform loading of $f$ is resumed, until some of the sites reach the threshold and the next avalanche starts. 

 The transmission parameter $\alpha$ measures the extent of non-conservation of the model. The system is conservative for $\alpha =0.25$, and is non-conservative for $\alpha <0.25$. A unit of time $t$ is taken to be the time required to load $f$ from zero to unity.

 In the OFC model,  boundary conditions play a crucial role. For example,  SOC state is realized under open boundary conditions, but is not realized under periodic boundary conditions.  The model under open boundary conditions goes into a special transient state where events of size 1 (single-site events) repeat periodically with period $1-4\alpha$  \cite{Middleton}. These single-site events occur in turn in a spatially random manner, but after time $1-4\alpha$, the same site topples repeatedly. Although such a periodic state consisting of single-site events is a steady state  under periodic boundary conditions, it is not a steady state under open boundary conditions because of the boundary. Indeed, clusters are formed near the boundary, within which the stress values are more or less uniform, and gradually invades the interior destroying the periodic state. Eventually, such clusters span the entire lattice, giving rise to an SOC-like steady state. Such clusters might be formed via synchronization between the interior sites and the boundary sites, the latter having a slower effective loading rate due to the boundary \cite{Middleton}. 

 In our simulation, the lattice studied contains $256\times 256$ sites with open boundary conditions, the pseudo-sequential updating proposed by Pinho {\it et al\/} being utilized \cite{Pinho}. Total of $2\times 10^9$ avalanches are generated where initial $10^8$ events are discarded as transients to ensure that the measurement is done in the steady state. 

 It has been realized that the OFC model exhibits an SOC-like critical or near-critical property. This is most clearly exemplified in a power-law-like size distribution of avalanches. As an example, we show in Fig.1 the size distribution of avalanches on a log-log plot for several values of the transmission parameter $\alpha$. The avalanche size $s$ is defined by the total number of ``topples'' in a given avalanche, which could be larger than the number of toppled sites because multi-toppling is possible in a given avalanche. In fact, it is observed that multi-toppling rarely occurs in the model except in the conservation limit or in the regime very close to it.

\begin{figure}[ht]
\begin{center}
\includegraphics[scale=0.7]{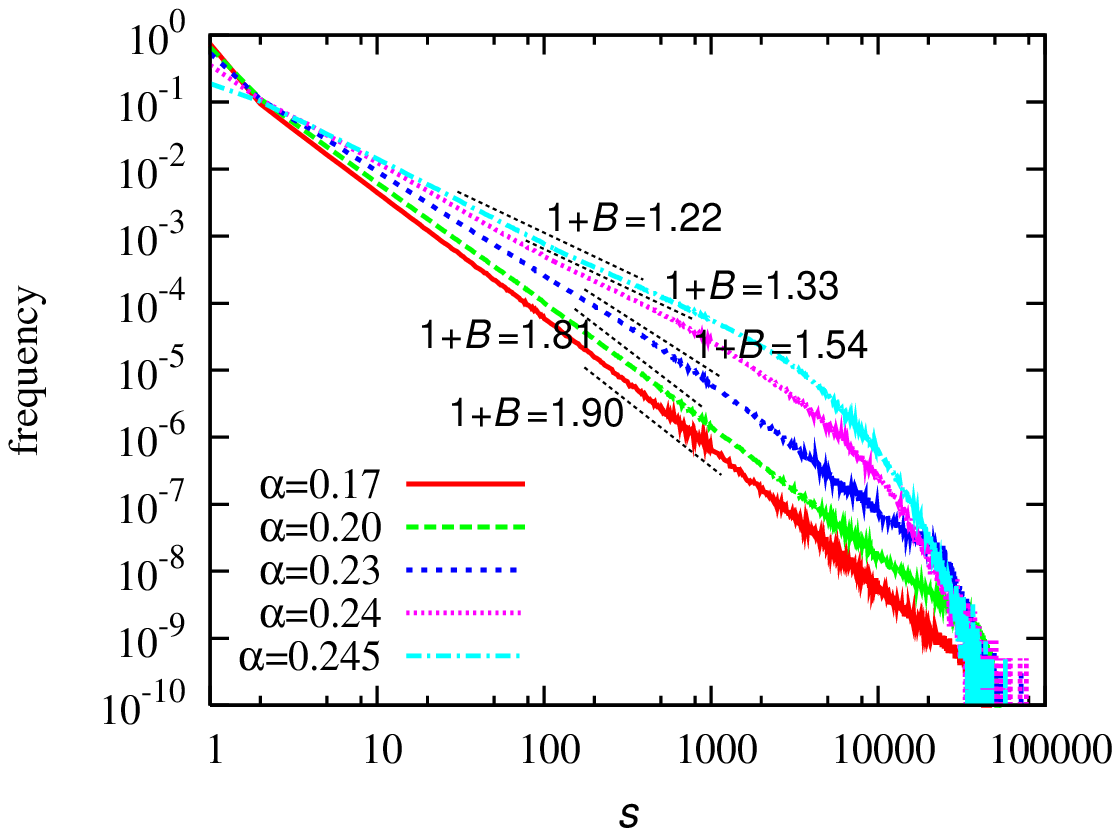}
\end{center}
\caption{
(Color online) The size distribution of seismic events (avalanches) for various values of the transmission parameter $\alpha $. The slope of the data gives the value of $1+B$, which is shown in the figure.
}
\end{figure}

 As can be seen from Fig.1, a straight-line behavior corresponding a power-law is observed consistently with the previous works.  The slope representing the $B$-value is not universal varying from $\simeq 0.90$ to $\simeq 0.22$ as $\alpha$ is varied from 0.17 to 0.245. The power-law feature is weakened as one approaches the conservation limit. The $B$-value is known to come around $2/3$ in real seismicity, while the $B$-value of the OFC model is distributed around this value \cite{OFC}.

 Hergarten {\it et al\/} observed that the OFC model also exhibited another well-known power-law feature of seismicity, {\it i.e.\/}, the Omori law (or the inverse Omori law) describing the time evolution of the frequency of aftershocks (foreshocks) \cite{Hergarten}. We show in Fig.2(a) on a log-log plot the frequency of aftershocks as a function of the time elapsed after the mainshock $t$. As a mainshock, we consider an event of its size greater than $s\geq s_c=100$. Again, a straight-line behavior corresponding a power-law is observed consistently with Ref.\cite{Hergarten}. Note that Hergarten {\it et al\/} imposed free boundary conditions, while we impose open boundary conditions here. The slope representing the Omori exponent $p$ is again not universal depending on the parameter $\alpha$ as $p=0.84$, 0.69 and 0.03 for $\alpha=0.17$,0.20 and 0.23, respectively. Since the $p$-value is known to come around unity in real seismicity, the $p$ value of the OFC model is not necessarily close to real observation. 

 Similar results are obtained also for foreshocks: See Fig.2(b). Here the $p$-value describing the inverse Omori law is equal to $p=0.71$, 0.50 and 0.01 for $\alpha=0.17$,0.20 and 0.23, respectively. As one approaches the conservation limit $\alpha=0.25$, both aftershocks and foreshocks tend to go away. Our results are consistent with the results reported by Hergarten {\it et al\/} \cite{Hergarten}.

\begin{figure}[ht]
\begin{center}
\includegraphics[scale=0.8]{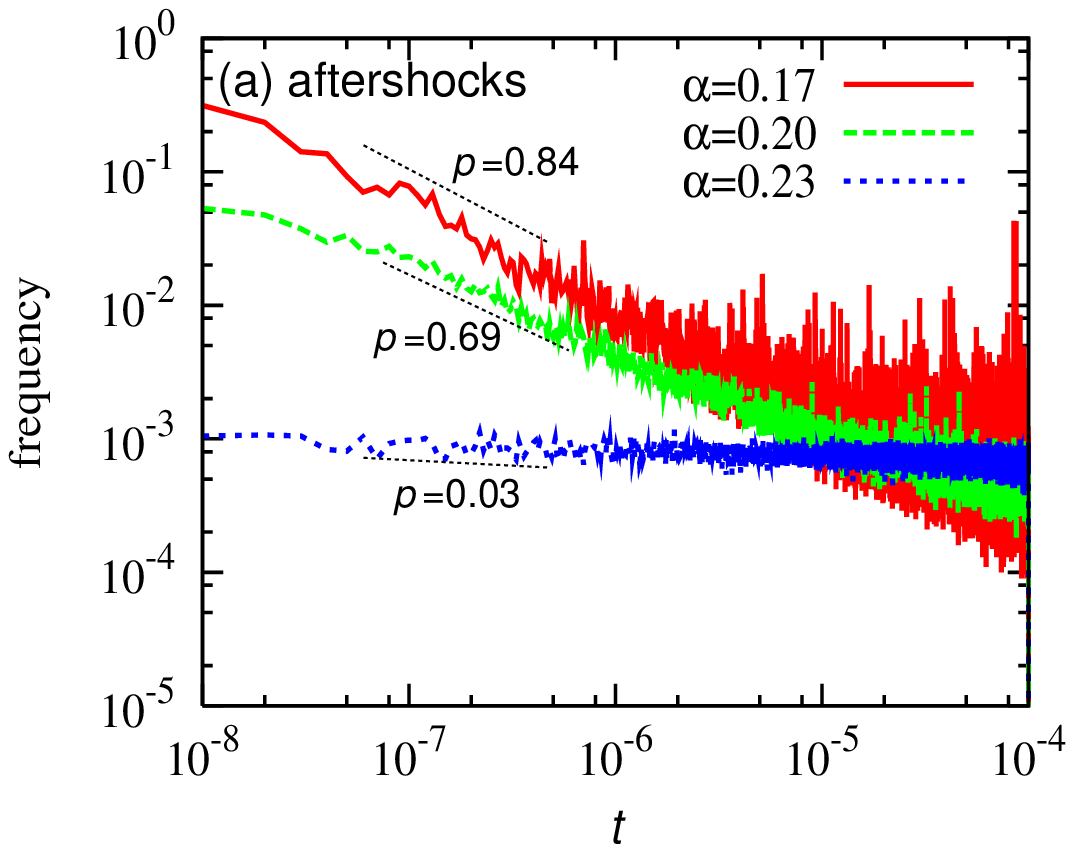}
\includegraphics[scale=0.8]{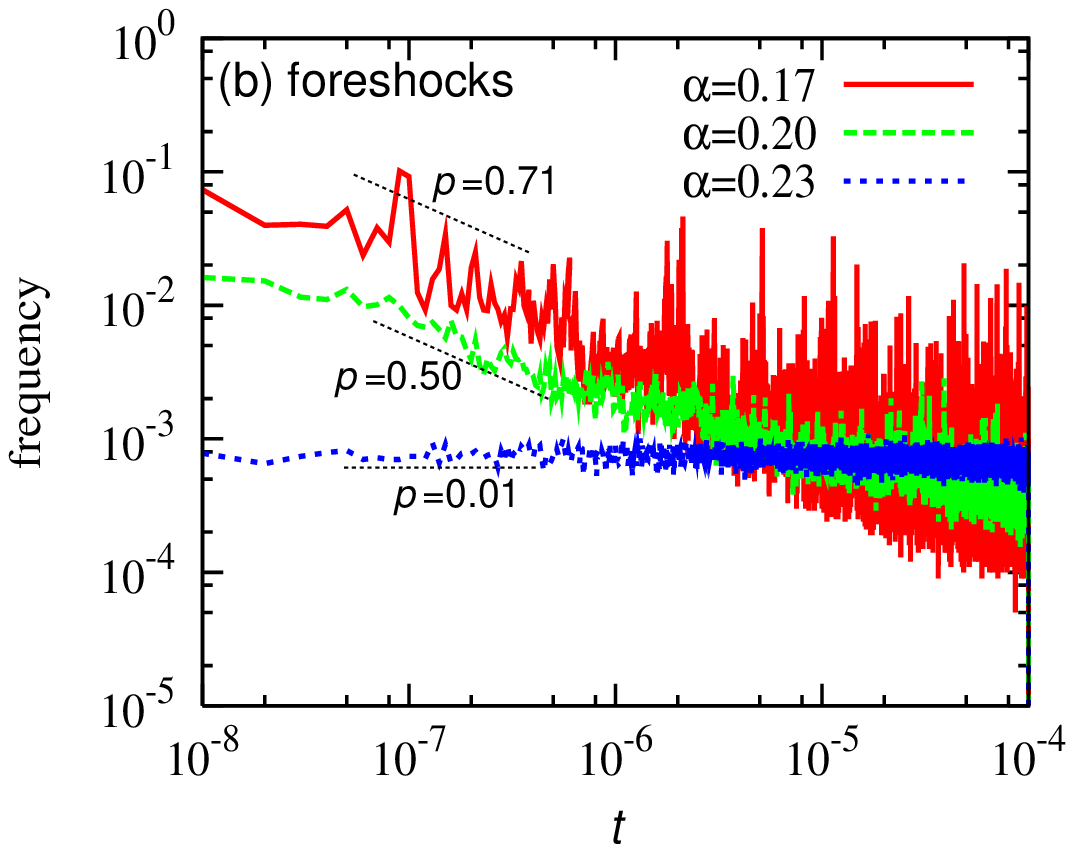}
\end{center}
\caption{
(Color online) The time dependence of the frequency of aftershocks (a), and of foreshocks (b), on a log-log plot for several values of the transmission parameter $\alpha$. Mainshocks are the events of their size greater than $s\geq s_c=100$. The time $t$ is measured with the occurrence of a mainshock as an origin. Aftershocks  and foreshocks are defined as events of arbitrary sizes which occur in the vicinity of mainshock with the range parameter $r_c=10$.
}
\end{figure}

 As such, the OFC model certainly possesses critical or near-critical properties like the GR law and the Omori law. By contrast, its {\it characteristic\/} properties, most notably revealed in the existence of ``asperity'' \cite{Kotani}, have received much less attention. In the next section, we investigate such characteristic properties of the OFC model, which apparently coexist with the critical properties mentioned in this section.

\section{Simulation results on the characteristic properties}

 In this section, we report on our simulation results on the local recurrence-time distribution and the stress distribution, which turn out to exhibit pronounced characteristic properties. 

 Let us begin with the definition of the {\it local\/} recurrence time.  In case of the globally defined recurrence time, the next avalanche to measure recurrence may occur anywhere on the entire lattice. In view of the ordinary sense of earthquake recurrence, however, it might be more natural to introduce the recurrence time $T$ and its distribution function $P(T)$ {\it locally\/}. Thus, one may define the local recurrence-time as the time passed until the next avalanche occurs with its epicenter lying in a vicinity of the preceding avalanche, say, within distance $r$ (in units of lattice spacing) of the epicenter of the preceding event, rather than in some remote place far away from the epicenter of the preceding event. One can also introduce the size threshold $s_c$ to look at the recurrence of large events of their size greater than $s\geq s_c$.

 The local recurrence-time distribution of the model, $P(T)$, was calculated in Ref.\cite{Kotani} with  varying the range parameter $r_c$.  The computed $P(T)$ exhibited a sharp $\delta$-function-like peak at $T=T^*=1-4\alpha$ which grew as $r_c$ got smaller, indicating that many (though not all) events of the OFC model repeated with a fixed  time-interval $T=T^*$. The peak position turned out to be independent of the range parameter $r_c$, the size threshold $s_c$, and the lattice size (as long as it was not too small). 

 In Fig.3, we show the local recurrence-time distribution $P(T)$ for avalanches whose size is greater than $s\geq s_c=100$ for fixed $r_c=10$, with varying the transmission parameter $\alpha$ toward the conservation limit $\alpha=0.25$. As $\alpha$ is increased toward $\alpha=0.25$, the $\delta$-function peak is gradually suppressed with keeping its position strictly at $T=1-4\alpha$. For $\alpha=0.245$, the $\delta$-function peak is no longer appreciable at the expected position $T=1-4\times 0.245$=0.02: See the arrow in the figure. Clearly, the $\delta$-function peak of $P(T)$ goes away toward the conservation limit $\alpha=0.25$.

\begin{figure}[ht]
\begin{center}
\includegraphics[scale=0.7]{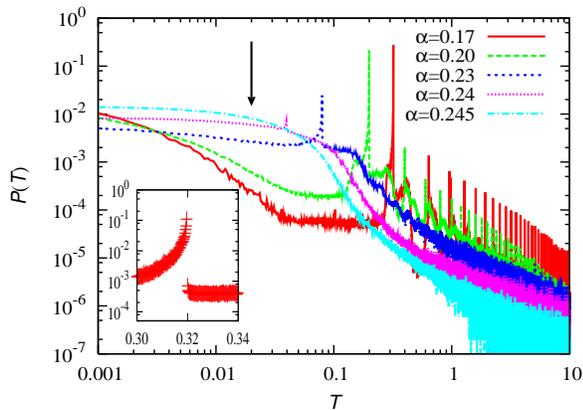}
\end{center}
\caption{
(Color online) Log-log plots of the local recurrence-time distributions of large avalanches of their size $s\geq s_c=100$  for a fixed range parameter $r_c=10$, with varying the transmission parameter $\alpha$. The arrow in the figure represents the expected peak position  for $\alpha=0.245$ corresponding to the period $T^*=1-4\alpha =0.02$. The inset is a magnified view of the main peak for the case of $\alpha=0.17$.
Log-log plots of the local recurrence-time distributions of large avalanches of their size $s\geq s_c=100$  for a fixed range parameter $r_c=10$, with varying the transmission parameter $\alpha$. The arrow in the figure represents the expected peak position  for $\alpha=0.245$ corresponding to the period $T^*=1-4\alpha =0.02$. The inset is a magnified view of the main peak for the case of $\alpha=0.17$.
}
\end{figure}

 As can be seen from Fig.3, in the longer time regime $T>T^*$, $P(T)$ exhibits behaviors close to power laws \cite{Kotani}, whereas the deviation from the power law becomes appreciable when one approaches the conservation limit. Hence, in earthquake recurrence of the model,  the characteristic or periodic feature, {\it i.e.\/}, a sharp peak in $P(T)$ at $T=T^*$, and the critical feature, {\it i.e.\/}, power-law-like behaviors of $P(T)$ at $T>T^*$, coexist.  Note that, while the peak at $T=T^*$ is sharp, it is not infinitely sharp with a finite intrinsic width as shown in the inset of Fig.3. The peak has an asymmetric shape with a tail on shorter-$T$ side.

 Furthermore, as was already noticed in Ref.\cite{Kotani}, the periodic events contributing to a sharp peak of $P(T)$ (``peak events'') possess a power-law-like size distribution very much similar to those observed for other aperiodic events. This is demonstrated in Fig.4, where the size distribution of the peak events is given for the case of $\alpha=0.23$, in comparison with the one for general events of $s\geq s_c=100$. (At this $\alpha$-value, $\alpha=0.23$, a slight excess is discernible at larger $s$ as was already noticed in Fig.1 above, both for all events and for the peak events.)  In that sense, the peak events certainly appear to be characteristic in time, but not necessarily so in their size.

\begin{figure}[ht]
\begin{center}
\includegraphics[scale=0.7]{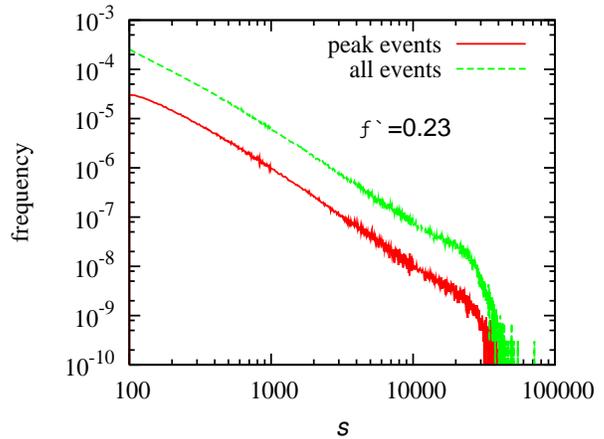}
\end{center}
\caption{
(Color online) The size distribution of the peak events as compared with the one of general events of $s\geq s_c=100$. The transmission parameter is  $\alpha=0.23$. The peak event is defined here as an event of its size greater than $s\geq s_c=100$ with its recurrence time belonging to the main peak of the local recurrence-time distribution function $P(T)$ given in Fig.3.
}
\end{figure}

 The fact that the peak events possess a power-law like broad distribution in their size $s$ also explains our observation of Fig.3 that the computed $P(T)$ exhibits a multi-peak structure consisting of sub-peaks located at multiples of $T^*=1-4\alpha$.  Reflecting such a size distribution of peaks events, many peak events occur with their size above our size threshold $s_c$ slightly. In such a situation, the ``next'' event occurring at $T=T^*$ after the first one might occur with its size slightly below the threshold $s_c$, and the ``second-next'' event occurring $T=2T^*$ after the first one might be counted as the next event in measuring the local recurrence time (remember that the peak in $P(T)$ has a small but finite width). In Ref.\cite{Kotani}, the major cause of the multi-peak structure of $P(T)$ was ascribed to the one associated with the range parameter $r_c$, but we now have found that the major cause of the multi-peak structure of $P(T)$ is the one associated with the size threshold $s_c$ rather than with the range parameter $r_c$: Refer also to our analysis of the asperity characteristics in \S 4.

 We have examined whether there exists appreciable difference between the peak events and the other events with regard to their aftershock/foreshock properties. However, both types of events accompany very much similar Omori-law-type aftershock/foreshock sequences.  Concerning their aftershock/foreshock sequences, we cannot identify any appreciable difference between the peak events and the other events.

 In Fig.5, we show for several $\alpha$-values the time-averaged stress distribution  in the steady state $D(f)$ averaged over all sites of the lattice. Though the system is loaded with a constant rate, the stress distribution is not uniform in the interval [0, 1]. Rather, for the case of $\alpha \leq 0.2$, $D(f)$ exhibits distinct steps. For $\alpha \leq 0.2$, these steps appear at appropriate multiples of the transmission parameter $\alpha$, {\it i.e.\/}, at $f=n\alpha$ for $f<1/2$ and at $1-n\alpha$ for $f>1/2$, with $n$ being an integer. For larger $\alpha>0.2$, no step seems to appear at multiples of $\alpha$, but $D(f)$ still exhibits a noticeable structure. In this sense, the stress distribution also exhibits a characteristic feature. Such a structure in $D(f)$ is likely to be originated from large-scale avalanches.

\begin{figure}[ht]
\begin{center}
\includegraphics[scale=1.0]{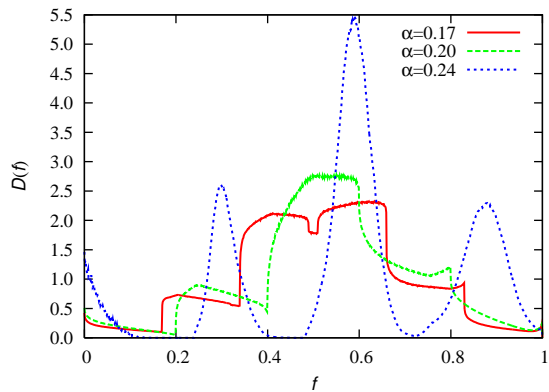}
\end{center}
\caption{
(Color online) The time-averaged stress distribution  in the steady state averaged over all sites of the lattice for several values of the transmission parameter $\alpha$. 
}
\end{figure}

\section{Simulation results on the asperity characteristics}

 In fact, the peak events, or near-periodic events, are closely related to the asperity-like phenomena \cite{Kotani}. The asperity represents a cluster of sites which ruptures simultaneously in a given avalanche, repeating many times with the same epicenter site and with almost the same period $T=T^*=1-4\alpha $ corresponding to the peak of the local recurrence-time distribution function $P(T)$. 

 We show in Fig.6  typical snapshots of the stress distribution immediately before and after an asperity event  for the case of $\alpha=0.2$. Within the asperity region, uniform stress state is formed before the rupture giving rise to a synchronized simultaneous rupture. Then, a discontinuous drop of the stress within a rupture zone (asperity) occurs before and after the avalanche. In fact, the same cluster except for a minor difference ruptures again after the time $T^*$. Such a rhythmic recurrence of rupture often repeats more than ten times. There is a clear tendency that such ``repeating times'' of the asperity sequence tend to be greater for smaller $\alpha$. In particular, the asperity-like phenomena themselves hardly exist near the conservation limit (remember that the local recurrence-time distribution $P(T)$ shown in Fig.3 no longer exhibits a peak at $\alpha=0.245$).

\begin{figure}[ht]
\begin{center}
\includegraphics[scale=0.75]{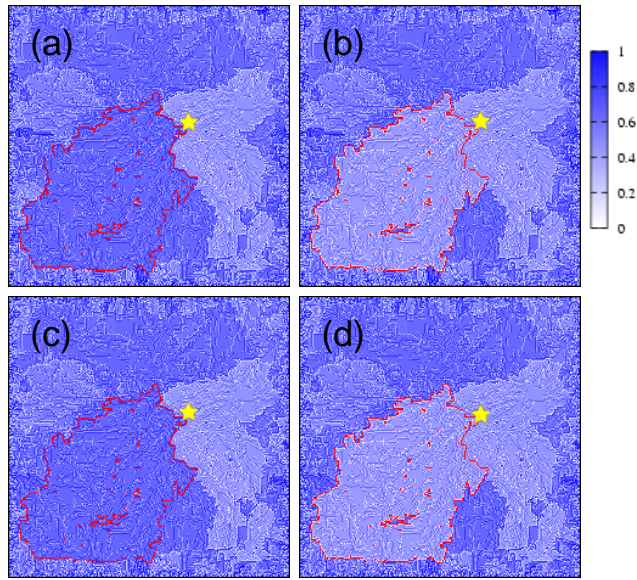}
\end{center}
\caption{
(Color online) Snapshots of the stress distribution for the case of $\alpha=0.2$; (a) immediately before a large event at time $t=t_0$, (b) immediately after this event, (c) immediately before the following event which occurs at time $t=t_0+T^* (T^*=0.2)$, and (d) immediately after this second event. Two events are of size $s=15891$ and $s=15910$ on a $L=256$ lattice. The region surrounded by red bold lines represents the rupture zone, while the star symbol represents an epicenter site which is located at the tip of the rupture zone.
}
\end{figure}

 The stress distribution in the asperity region tends to be ``discretized'' to certain values. In Fig.7(a) and (b), we show for the case of $\alpha=0.17$ the stress distribution $D(f)$ of the asperity sites immediately before (a) and after (b) an avalanche, averaged over asperity events. We define here the asperity as a ruptured cluster of its size $s\geq s_c=100$ belonging to the peak event of the local recurrence-time distribution function.  As can be seen from the figure, $D(f)$ now consists of several ``spikes'' located at appropriate multiples of the transmission parameter $\alpha$, {\it i.e.\/},  at $1-n\alpha$ before the rupture and at $f=n\alpha$ after the rupture, with $n$ being an integer. One can now see that the structure found in the time-averaged stress distribution function shown in Fig.5 is originated from those spikes associated with the asperity event.

\begin{figure}[ht]
\begin{center}
\includegraphics[scale=0.7]{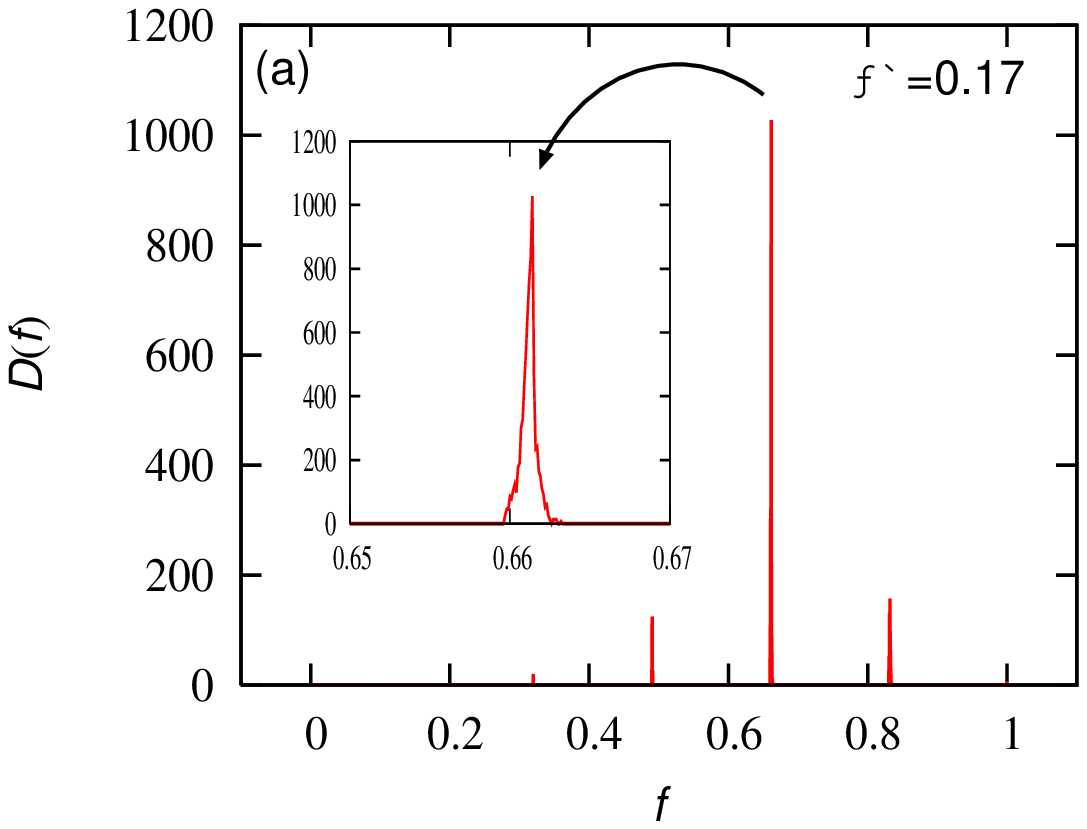}
\includegraphics[scale=0.7]{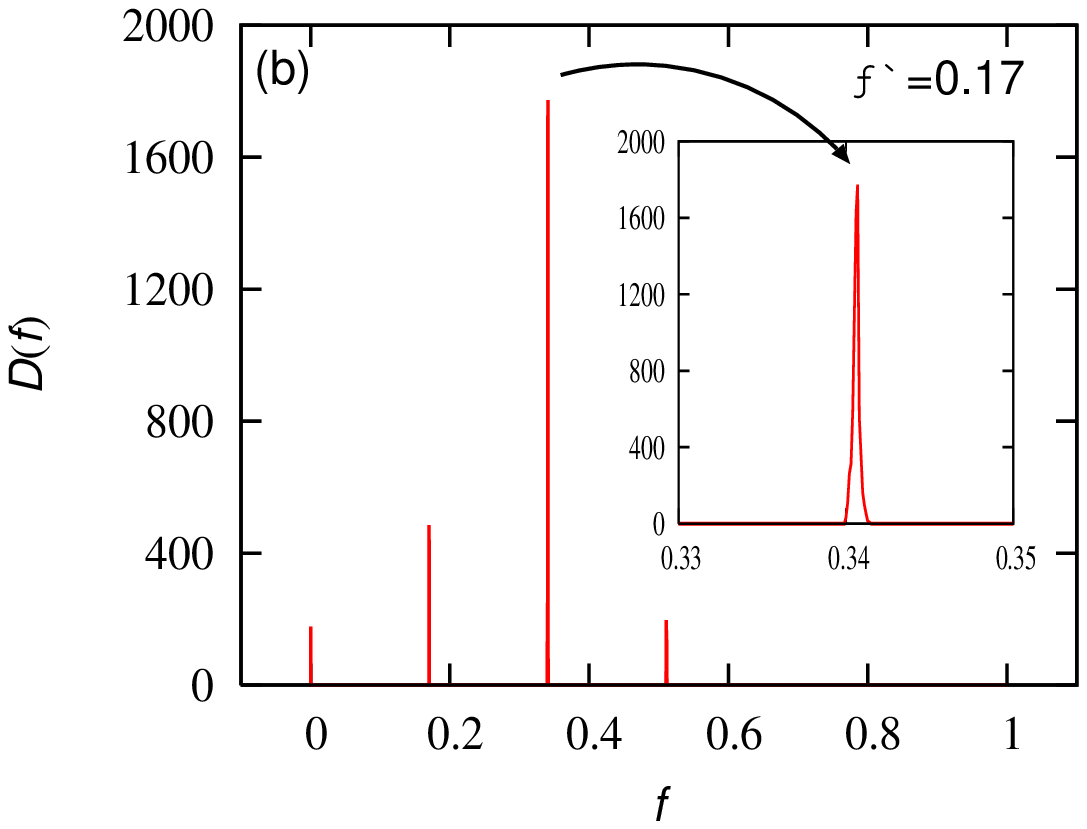}
\end{center}
\caption{
(Color online) The stress distribution $D(f)$ of each site contained in the rupture zone of the asperity event, just before (a) and after (b) the asperity event. An asperity event is defined here as an event of its size greater than $s\geq s_c=100$ belonging to the main peak of the local recurrence-time distribution function. The transmission parameter is $\alpha=0.17$. The inset is a magnified view of the main peak.
}
\end{figure}

 In fact, such a tendency of ``quantization'' or ``spikes'' also exists in the non-asperity events as well, but in a less pronounced manner. In Fig.8, we show the stress distribution at the time of toppling of each site contained in the rupture zone for both cases of the asperity events and the non-asperity events. The stress at the time of toppling generally exceeds the threshold value $f_c=1$ (except for an epicenter site where it is identically unity). As can be seen from Fig.8, the stress value at the time of toppling is more concentrated on the threshold value $f_c=1$ for the asperity events than that for the non-asperity events, the latter exhibiting a broader tail toward larger values of $f>1$.

\begin{figure}[ht]
\begin{center}
\includegraphics[scale=0.7]{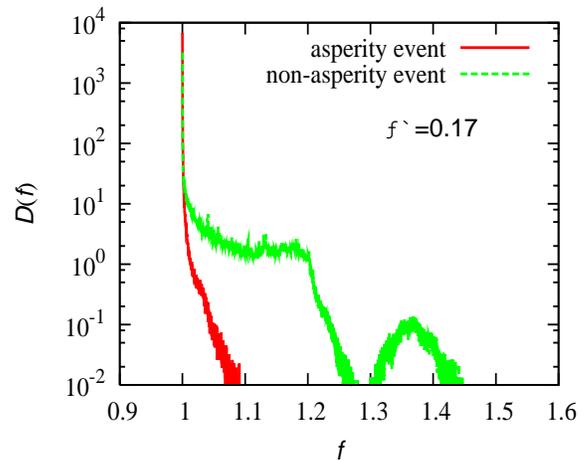}
\end{center}
\caption{
(Color online) The stress distribution $D(f)$ at the time of toppling of each site contained in the rupture zone of large events with $s\geq s_c=100$. The data are given for both cases of the asperity events and the non-asperity events. An asperity event is defined here as an event of its size greater than $s\geq s_c=100$ belonging to the main peak of the local recurrence-time distribution function. The transmission parameter is $\alpha=0.17$. 
}
\end{figure}

 Furthermore, as the asperity events repeat, the tendency of the stress concentration is more and more enhanced. In Fig.9, we show the time sequence of the stress distribution at the time of toppling for the asperity events. As the asperity events repeat, the stress distribution tends to be narrower, being more concentrated on the threshold value $f_c=1$. In fact, one can show that the stress distribution at the time of toppling tends to be more concentrated on the threshold value $f_c=1$ as the asperity events repeat. Namely, once each site starts to topple more or less at similar stress values close to the threshold value $f_c=1$, this tendency is more and more evolved as the asperity events repeat. The stress concentration tends to be self-organized. In such a situation, all sites in an asperity cluster (except for an epicenter site) possess the stress value close to $1-\alpha$ just before the event starts. Assume that an arbitrary site involved in the interior of the asperity cluster possesses the stress value $f=1-\alpha +x$ with $x << 1$, just before the event starts. One can show that, when this event is over and the interseismic period of $T^*=1-4\alpha$ has passed, the stress value at this site becomes $1-\alpha +\delta^* x$ with $\delta^*$ being less than unity under rather general conditions, as long as the stress values at neighboring sites  just before the event are also close to $1-\alpha$.  Details of the derivation are given in Appendix.

\begin{figure}[ht]
\begin{center}
\includegraphics[scale=0.7]{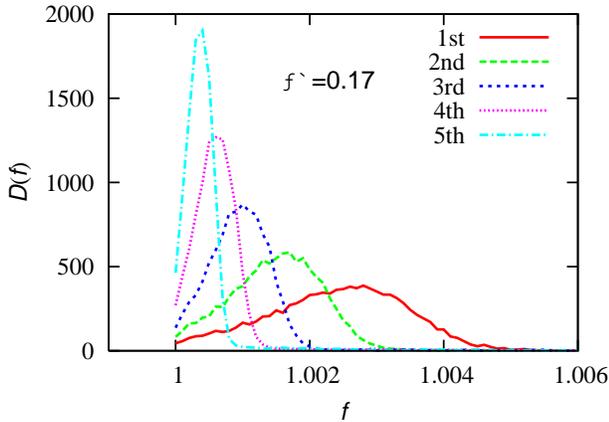}
\end{center}
\caption{
(Color online) The 
 time sequence of the stress distribution $D(f)$ at the time of toppling of each site contained in the rupture zone of the asperity events. An asperity event is defined here as an event of its size greater than $s\geq s_c=100$ belonging to the main peak of the local recurrence-time distribution function. The transmission parameter is $\alpha=0.17$. As the events repeat, the stress distribution at the time of toppling gets more and more concentrated on the borderline value $f_c=1$.
}
\end{figure}

 We note that, if all sites topple at the stress value close to the threshold  $f_c= 1$ in the asperity events, it immediately explains our observation that the period of the asperity events, corresponding to the main peak of the local recurrence-time distribution, is equal to $T=T^*=1-4\alpha$. To see this, one only needs to remember the conservation law of the stress, {\it i.e.\/}, the stress dissipated at the time of toppling, which is $1-4\alpha$ per site if the toppling occurs exactly at $f=1$, should match the stress loaded during the interval time $T$.

\begin{figure}[ht]
\begin{center}
\includegraphics[scale=0.75]{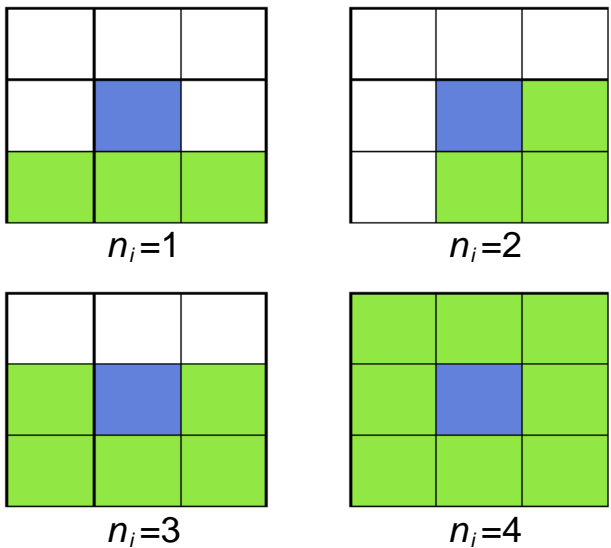}
\end{center}
\caption{
(Color online) Four types of epicenter sites at the center of the figure (blue). Shaded region represents the rupture zone. 
}
\end{figure}

\begin{figure}[ht]
\begin{center}
\includegraphics[scale=0.8]{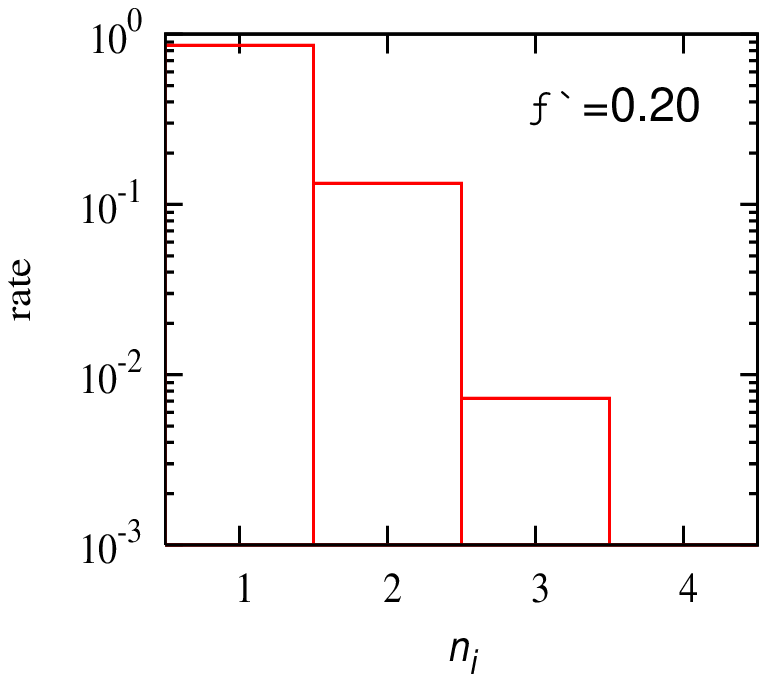}
\includegraphics[scale=0.8]{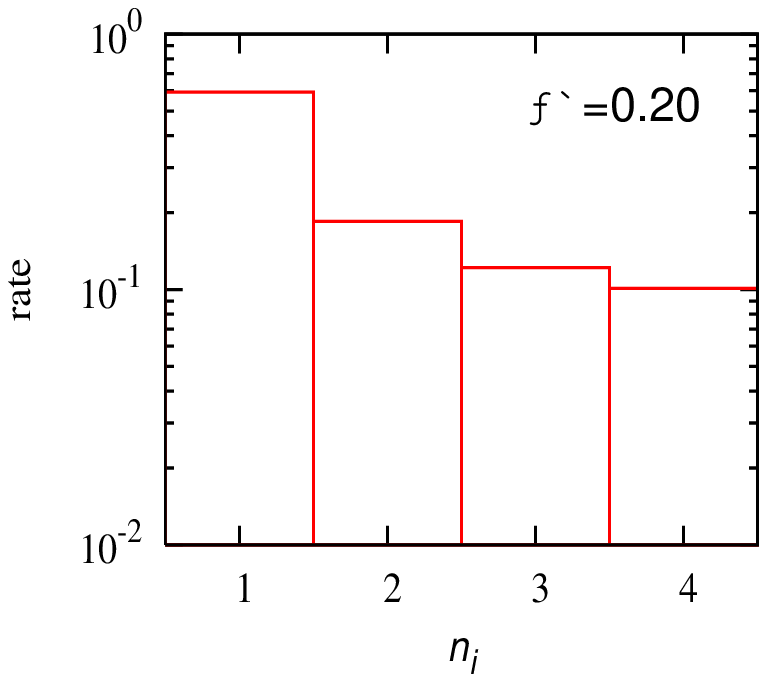}
\end{center}
\caption{
(Color online) The rate of the type of epicenter sites, for the asperity events (a) and for other non-asperity events (b), plotted on a semi-logarithmic scale. The abscissa $n_i$ represents the number of nearest-neighbor sites contained in the rupture zone: See Fig.10.  The transmission parameter is $\alpha=0.20$.
}
\end{figure}

 Next, we examine where in the rupture zone an epicenter site is located for both cases of the asperity event and for the non-asperity event. We classify the site $i$ into four types according to the number $n_i$ of its nearest-neighbor sites which topple during a given avalanche. The cases $n_i$=1,2,3 and 4 typically represent an epicenter site lying at the tip, corner, boundary and interior of the rupture zone, respectively: See Fig.10. In Fig.11, we show  on a semi-logarithmic plot the rate of each type of an epicenter site of large events with  $s\geq s_c=100$, for the asperity events (a) and for the non-asperity events (b). The transmission parameter is $\alpha=0.20$. As can be seen from Fig.11, there is a pronounced tendency that the epicenter site is located at smaller $n_i$, particularly at $n_i=1$ corresponding to the tip of the rupture zone. This tendency is more enhanced in the asperity events than in other events. It turns out that the interior site with $n_i$=4 can never be an epicenter of the asperity events while a small number of such events ($\sim 10$\%) exist for the non-asperity event. 

 Clear difference exists between the asperity and the non-asperity events in other quantities as well. In Fig.12, we show  for the case of $\alpha=0.2$ the temporal variation  of the frequency of both the asperity events and the non-asperity events observed in a typical run of a 512$\times $ 512 lattice. The frequency is defined here as the number of each type of events occurring in a time bin of one $T^*$ ($=1-4\alpha=0.2$).  As can be seen from the figure, the frequency of the non-asperity events fluctuates around a constant value, whereas that of the asperity events exhibits much wilder temporal variation, occasionally vanishing altogether. In other words, there appear in turn an active period and a calm period in the asperity events, while no such distinction seems to exist in the non-asperity events.

\begin{figure}[ht]
\begin{center}
\includegraphics[scale=1.1]{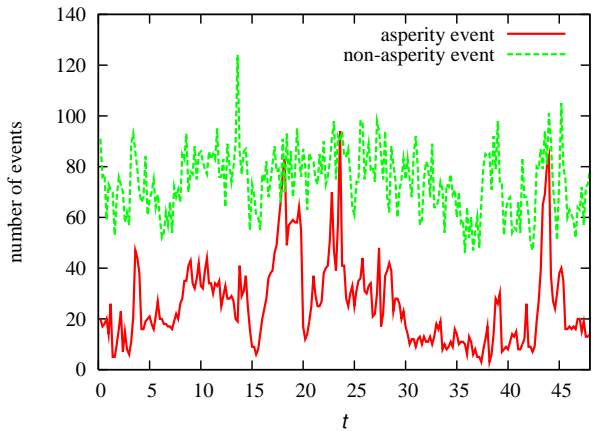}
\end{center}
\caption{
(Color online) The time variation  of the frequency of the asperity events and of the non-asperity events in a typical run of a 512$\times $ 512 lattice. The transmission parameter is $\alpha=0.2$. The frequency is defined by the number of each type of events which occur in a time bin of one $T^*$ ($=1-4\alpha=0.2$). 
}
\end{figure}

\begin{figure}[ht]
\begin{center}
\includegraphics[scale=1.15]{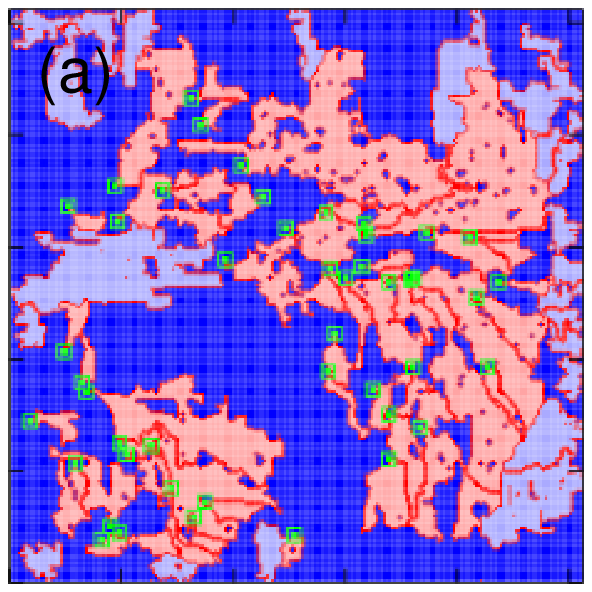}
\includegraphics[scale=1.15]{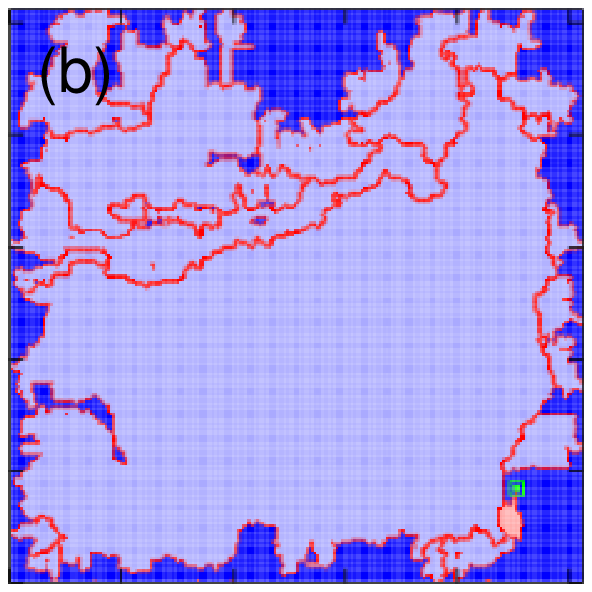}
\end{center}
\caption{
(Color online) The spatial distribution of large events in a typical run of a 256$\times $256 lattice. The rupture zone of both the asperity events and  the non-asperity events, which occur in a time period of one $T^*$, are shown. Fig.(a) corresponds to the midst of the ``asperity period'' where the frequency of the asperity events takes its maximum, while Fig.(b) corresponds to the midst of the ``non-asperity period'' where the frequency of the asperity events takes its minimum. The transmission parameter is $\alpha=0.2$. Dark blue area represents either only small events of its size $s < s_c=100$ or no event occurring during the time interval of $T^*$. Thin orange (thick gray) area represents asperity (non-asperity) cluster of its size  $s > s_c=100$. The red bold line represents the rupture zone of each large event, while green square represents an  epicenter site of the asperity event. 
}
\end{figure}

 The asperity events tend to cluster not only in time but also in space. In Fig.13, we show for the case of $\alpha=0.2$ typical spatial distributions of both the asperity events and the non-asperity events on a 256$\times $256 lattice. The rupture zone of both types of events, which occur in a time period of one $T^*$, are shown together with the epicenter site of the asperity events. Fig.(a) corresponds to the midst of the ``asperity period'' where the frequency of the asperity events takes its maximum, while Fig.(b) corresponds to the midst of the ``non-asperity period'' where the frequency of the asperity events takes its minimum.  As can be seen from Fig.13(a), the asperity events tend to cluster spatially: In particular, the epicenter site of the asperity event, which tends to lie at the tip or at the corner of the rupture zone, is contiguous either to the rupture zone of other asperity events or to the area of no (small) events, but not contiguous to the rupture zone of non-asperity events. Since the multi-toppling rarely occurs in the model, this observation means that the stress the epicenter site of the asperity event receives from its neighboring sites tends to be nearly equal to $\alpha$. There also exists a tendency that the asperity cluster lies in the interior of the lattice rather than near the boundary, as was already noticed  in Ref.\cite{Kotani}. 

 In the ``non-asperity period'', by contrast, large events are dominated almost exclusively  by the non-asperity events: See Fig.13(b). Since the number of the non-asperity events is more or less kept constant in time as can be seen from Fig.12, the size of the non-asperity cluster tends to get larger in the non-asperity period, which seems consistent with the pattern observed in Fig.13(b).

 In the following, we shall concentrate on the asperity events. An epicenter site topples at the beginning of a given event releasing its stress to zero. If the epicenter site $i$ is of the type $n_i$ ($n_i=1,2,3,4$), it receives the stress from its nearest-neighbor sites $n_i$ times during this asperity event. Since any toppling in the asperity event occurs close to the borderline stress $f_c=1$, the stress value the epicenter site receives during this event is equal to $n_i \alpha$. As mentioned, in asperity events, the same site $i$ is likely to be an epicenter of the next event again, {\it i.e.\/}, reaches the threshold stress $f_c=1$ earlier than any other site in the rupture zone of this event. This might sound a bit surprising at first, since, as shown in Fig.11, an epicenter site tends to be located at the tip position of the rupture zone with $n_i=1$, which means that the epicenter site is in a low stress state just after the preceding event. However, such an epicenter site rapidly develops its stress not only by stress loading but also by receiving stress via the topplings of its $4-n_i$ nearest-neighbor sites not belonging to the rupture zone of the preceding event. Here recall that the stress transfer  the epicenter site $i$ receives from the nearest-neighbor sites during the interval time tend to be nearly equal to $\alpha$.

 Then, one can see the reason why the same epicenter site becomes an epicenter of the next asperity event again. The stress the epicenter site $i$ receives from the one-site events (or other asperity events) during the interval time is equal to $(4-n_i)\alpha$ so that the time needed for the site $i$ to reach the threshold stress $f_c=1$ is $T=1-n_i\alpha-(4-n_i)\alpha=1-4\alpha=T^*$, which is just the interval time of the asperity events observed in the local recurrence-time distribution. 

 Next, let us consider the stress at an arbitrary site $i$ in the rupture zone other than the epicenter site.  We suppose that, during a given asperity event, $m_i$ among $n_i$ nearest-neighbor sites topple before the site $i$ and the remaining $n_i-m_i$  nearest-neighbor sites topple after the site $i$. By definition, $1\leq m_i\leq n_i$. Then, the stress at the site $i$ just after the asperity event is $(n_i-m_i)\alpha$. The stress the site $i$ receives during the interval time from the sites not belonging to the rupture zone of the preceding event is $(4-n_i)\alpha$. Hence, the time needed for the site $i$ to reach the threshold stress $f_c=1$ is $T=1-(n_i-m_i)\alpha-(4-n_i)\alpha=1-4\alpha+m_i\alpha$, which is always greater than the corresponding time for the epicenter site estimated above, $1-4\alpha$, since $m_i\geq 1$. Thus, an epicenter site of the preceding asperity event tends to be an epicenter of the next asperity event again.

 One sees that the value of $n_i$ does not matter in the proof given above. Therefore, our observation in Fig.11 that a tip (or a corner) site tends to be an epicenter of the avalanche can not be explained solely from the above argument. At the moment, we do not know the reason why an epicenter site tends to lie at the tip or the corner of the rupture zone rather than in its interior in the OFC model, though the tendency is quite distinct in our numerical simulation.

\begin{figure}[ht]
\begin{center}
\includegraphics[scale=0.8]{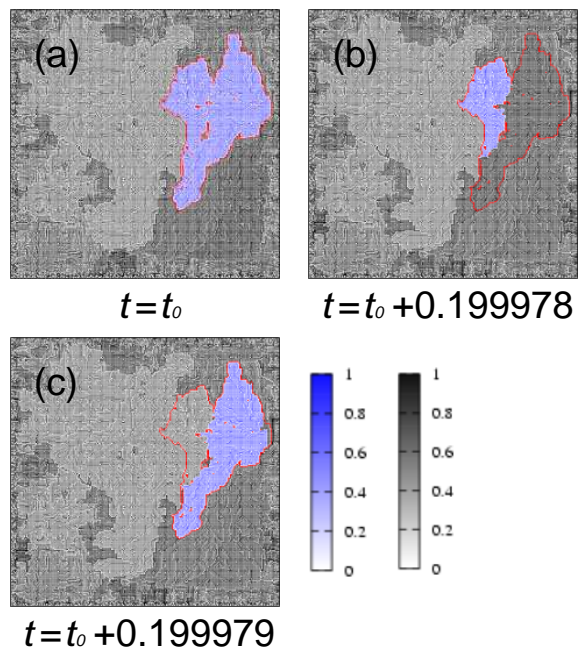}
\end{center}
\caption{
(Color online) A typical pattern of the asperity-sequence interruption, where one big asperity is divided into two parts. The transmission parameter is $\alpha=0.2$.  Fig.(a), (b) and (c) correspond to the time $t=t_0$ (a), $t=t_0+0.199978$ (b) and $t=t_0+0.199979$ (c).  The region surrounded by the red bold line represents the rupture zone of the first event.
}
\end{figure}

 The asperity events repeat over many times, but they do not last forever. After all, our model is a spatially uniform model so that no interior site can be special in the long-time limit. Then, we also study the manner how the asperity sequence is interrupted.  For the asperity sequence interruption, there seem to be two patterns. The first pattern is a detachment of the asperity. Namely, one big asperity is divided into two parts, each of which collapse not at the same time, but at different times, though still mutually close in time. An example of this pattern is given in Fig.14. The second pattern is an enlargement of the asperity. Namely, a given asperity event gets involved in other larger event becoming a part of it, which can occur when the asperity fails to rupture at the regular period $T^*=1-4\alpha$ due to some reason. An example of the latter pattern is given in Fig.15.

\begin{figure}[ht]
\begin{center}
\includegraphics[scale=0.8]{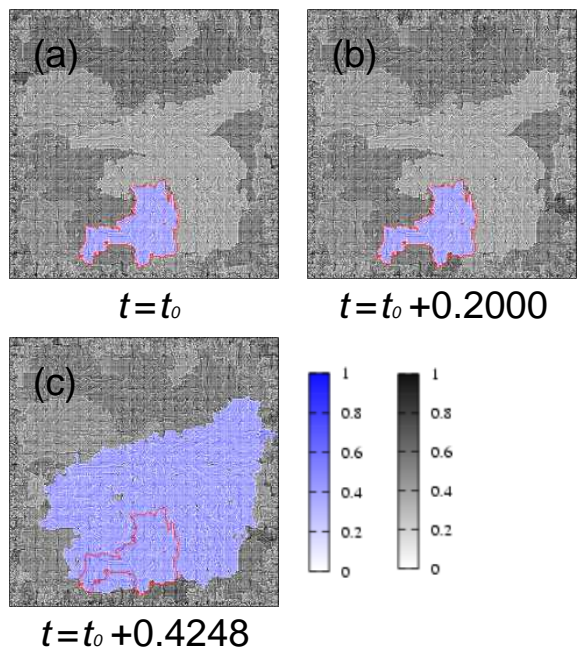}
\end{center}
\caption{
(Color online) Another typical pattern of the asperity-sequence interruption, where a given asperity event gets involved in other larger event becoming a part of it.  The transmission parameter is $\alpha=0.2$. Fig.(a), (b) and (c) correspond to the time $t=t_0$ (a), $t=t_0+0.2000$ (b) and $t=t_0+0.4248$ (c). The region surrounded by the red bold line represents the rupture zone of the first event.
}
\end{figure}

\section{Summary and discussion}

 We studied the properties of the OFC model of earthquakes. In this model, characteristic and critical properties coexist in an intriguing manner. We computed the magnitude distribution, the recurrence-time distribution, the stress distribution, and investigated various asperity characteristics in some detail. Interestingly, the OFC model exhibits not only well-known critical features like the GR law and the Omori law, but also pronounced characteristic or periodic features, most notably exemplified in the occurrence of asperity-like phenomena. We have found that the stress dissipation, {\it i.e.\/}, the non-conservation of the stress, is essential for the asperity-like phenomena to occur. We have also shown that a key ingredient in the asperity formation is a self-organization of the highly concentrated stress state. Such a stress concentration immediately explains why the interval time of the asperity events is equal to $1-4\alpha$, and why the same site becomes an epicenter in the asperity sequence. The asperity-like events in the OFC model closely resemble those familiar in seismology \cite{Scholzbook}, in the sense that almost the same spatial region ruptures repeatedly with a common epicenter site and  with a common period. 

 Indeed, apparent coexistence of critical and characteristic features is a notable feature of real earthquake phenomena as well. The critical features are most typically seen  in various power-laws observed in statistical properties of earthquakes, {\it e.g.\/}, the GR law and the Omori law as described. Typical examples of characteristic features of real earthquakes may be, {\it e.g.\/}, the asperities observed along the subduction zone in northeastern Japan, particularly repeating earthquakes off Kamaishi \cite{YamanakaKikuchi,Matsuzawa}. In seismology, the concept of earthquake cycles has been used in long-term probabilistic earthquake forecasts \cite{Scholzbook,Nishenko,WG,Kato}.

 Thus, the OFC model, though an extremely simplified model, may capture some of the essential ingredients necessary to understand apparent coexistence of critical and characteristic properties in real earthquakes. In seismology, the origin of the asperity is usually ascribed to possible inhomogeneity in the material property of the crust or in the external conditions. We wish to stress here, however, that in the present OFC model there is no built-in inhomogeneity in the model parameters nor in the external conditions, yet the ``asperity'' is self-generated from the spatially uniform evolution-rule and the model parameter. 

 As mentioned, the asperity in the OFC model is not a permanent one. After all, the model is spatially uniform. Meanwhile, we have observed that the asperity events tend to cluster both in time and in space. In particular, the asperity events exhibit ``active'' and ``calm'' periods in turn. In its active period, the asperity often exists stably over many earthquake recurrences. Thus, one needs to keep in mind a possibility that, even in the asperity formation of real seismicity, inhomogeneity might play only a secondary role.

 Of course, real crust is certainly inhomogeneous, and it is important to clarify how such inhomogeneity affects, or does not affect, the critical and the characteristic properties of earthquakes. In this context, it might be interesting to study the corresponding properties of the {\it inhomogeneous\/} OFC model in comparison with those of the present homogeneous OFC model. Among others, the question whether the asperity-like phenomena and the associated characteristic properties would survive the build-in inhomogeneity or not is of particular interest.  The answer to this question seems to be ``yes'', although some of the properties change from the ones observed for the present homogeneous model and its details often depend on the specific parameter the model. The properties of such inhomogeneous OFC model will be reported in our forthcoming publication \cite{Yamamoto}.

The authors are thankful to Prof. N. Kato for useful discussion. This study was supported by Grant-in-Aid for Scientific Research 19052006. We thank ISSP, Tokyo University for providing us with the CPU time.

\begin{center}
    {\bf APPENDIX }
\end{center}

In this appendix, we show for the asperity event that the stress distribution at the time of toppling tends to concentrate more on the threshold value $f_c=1$ as the asperity events repeat. 
 
 First, we look at the site $j$ which is the nearest neighbor of an epicenter site 0.  Let the stress of the site $j$ be $f_j=1-\alpha+x$. We assume $0<x<<1$, which is equivalent to assuming that this site topples nearly at its threshold value $f_c=1$. Among the four nearest-neighbor sites of the site $j$, one is an epicenter 0.   The epicenter site always topples at $f=f_c=1$ and distribute the stress $\alpha$ to its nearest-neighbor sites including $j$. Let the stress of other three nearest-neighbor sites of the site $j$ be $f_i=1-\alpha+x+\delta_i x$ ($i=1,2,3$): See Fig.16. After the toppling of the epicenter site 0, the site $j$ reaches the stress $1+x>1$ and topples. The three nearest-neighbor sites of the site $j$ then reach the stress $f_i+\alpha f_j = 1+x+\alpha x+\delta_i x$ ($i=1,2,3$) and topple. The site $j$ then receives the stress from the sites 1, 2, and 3, reaching the stress $3\alpha(1+x+\alpha x)+\alpha(\delta_1+\delta_2+\delta_3)x$. As long as the site $j$ does not topple again in this asperity event, which is usually the case, this is the stress value at the site $j$ immediately after this asperity event. After the interseismic period $T^*=1-4\alpha$, the stress at the site $j$ becomes $1-\alpha+3\alpha(1+\alpha +\delta )x$ with $\delta\equiv (\delta_1+\delta_2+\delta_3)/3$. Then, the condition of the stress concentration occurring at the site $j$ is $3\alpha(1+\alpha +\delta )<1$. Namely, as the asperity events repeat, the stress at the time of toppling gradually approaches the threshold value $f_c=1$ if the condition $\delta < \delta_c = \frac{1}{3\alpha} - (1+\alpha)$ is fulfilled.

\begin{figure}[ht]
\begin{center}
\includegraphics[scale=0.25]{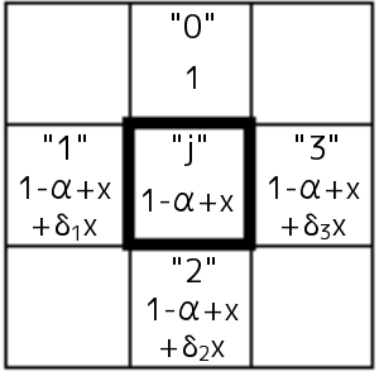}
\end{center}
\caption{
Illustration of neighboring toppling sites in the rupture zone, with its initial stress values. 
}
\end{figure}

 Now, we extend the above discussion to more general situation where, among the four nearest-neighbor sites of the site $j$, (i) one site 0 topples before the site $j$ and three other sites 1,2 and 3 topple after the site $j$. Here we consider the site $j$ to be an interior site with all its nearest-neighbor sites contained in the rupture zone of the asperity event. Let the stress of the site $j$ be $f_j=1-\alpha+x$ as above, the stress of the site toppling before $j$ be $f_0=1+x+\delta_0 x$, and the ones after $j$ be $f_i=1-\alpha +x+\delta_i x$ ($i=1,2,3$). After the toppling of the site $0$, the site $j$ reaches the stress $1+x+\alpha (1+\delta_0)x$ and topples. The three neighboring sites 1, 2 and 3 then reach the stress $f_i+\alpha f_j = 1+x+\alpha x+\delta_i x+\alpha^2(1+\delta_0)x$ ($i=1,2,3$) and topple. The site $j$ then receive the stress from the sites 1,2, and 3, reaching the stress $3\alpha(1+x+\alpha x+\alpha^2 x)+3\alpha\delta x + 3\alpha^3 \delta_0 x$. As long as the site $j$ does not topple again in this asperity event, which is usually the case, this is the stress value at the site $j$ just after this asperity event. After the period $T^*=1-4\alpha$, the stress at the site $j$ becomes $1-\alpha+3\alpha(1+\alpha +\alpha^2+\delta +\alpha^2\delta_0)x$.  Then, the condition of the stress concentration occurring at the site $j$ is $\delta <  \delta_c = \frac{1}{3\alpha} - \{1+\alpha +\alpha^2(1+\delta')\}$ where $\delta'=\delta_0$.

\begin{figure}[ht]
\begin{center}
\includegraphics[scale=1.3]{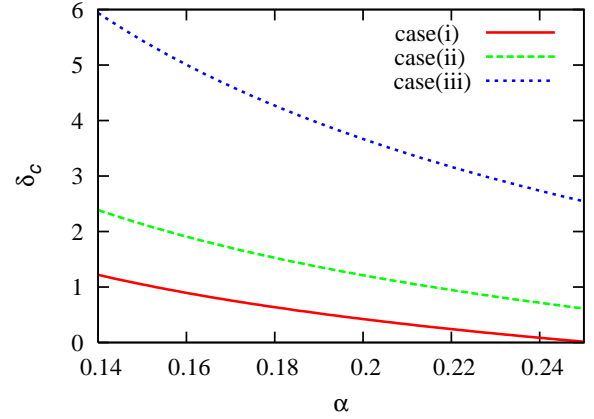}
\end{center}
\caption{
(Color online) The threshold value for the stress concentration, $\delta_c$, plotted versus the transmission parameter $\alpha$ in the case of $\delta'=0.1$. The three processes (i) $\sim$ (iii) are described in the text. One can see that the stress concentration process, expected at $\delta < \delta_c$, is readily operative in most cases, since $\delta_c$ is large except for the process (i) close to the conservation limit $\alpha=0.25$.
}
\end{figure}

 One can repeat similar discussion to the other cases where, among the four nearest-neighbor sites of the site $j$, (ii) two sites  0 and 1 topple before the site $j$ and the two other sites 2 and 3 topple after the site $j$, (iii) three sites 0, 1 and 2 topple before the site $j$ and one other site 3 topples after the site $j$, and (iv) all four sites topple after the site $j$. In the case (ii), the condition of the stress concentration is $\delta <  \delta_c = \frac{1}{2\alpha} - \{1+\alpha +2\alpha^2(1+\delta')\}$ with $\delta =(\delta_2+\delta_3)/2$ and $\delta' =(\delta_0+\delta_1)/2$. In the case (iii), the condition of the stress concentration is $\delta <  \delta_c = \frac{1}{\alpha} - \{1+\alpha +3\alpha^2(1+\delta')\}$ with $\delta =\delta_3$ and $\delta' =(\delta_0+\delta_1+\delta_2)/3$. The case (iv) is exceptional, where the stress of the site $j$ after the event is always zero. The $\alpha$-dependence of $\delta_c$ is shown in Fig.17 in the above cases of (i) $\sim$ (iii) with $\delta'=0.1$. In fact, the $\delta_c$-value is rather insensitive to the $\delta'$-value, since, in the expression of $\delta_c$, the coefficient of the $\delta'$-term is smaller than the leading term by factor of $O(\alpha^3)$ with $\alpha<0.25$.
 
 The above analysis indicates that, once each site starts to topple more or less at similar stress values close to the threshold value $f_c=1$, this tendency is more and more evolved as the asperity events repeat, {\it i.e.\/}, the stress concentration tends to be self-organized.

\end{document}